\documentstyle[12pt,aasms4]{article}
\newcommand{\doublespacing}{\let\CS=\@currsize\renewcommand{\baselinestretch}
{1.0}\tiny\CS}

\newcommand {\beq} {\begin{equation}}

\newcommand {\eeq} {\end{equation}}

\begin{document}

\title{Variation  of Physical constants, Redshift and the Arrow of time } 
\author{Menas Kafatos\altaffilmark{1}, Sisir Roy\altaffilmark{1,2} and Malabika Roy\altaffilmark{1}}
\centerline{$^1$Center for Earth Observing and Space Research, School of  Computational Sciences} 
\centerline{George Mason University, Fairfax, VA  22030  USA}
\centerline{$^2$Physics and Applied Mathematics Unit, Indian Statistical Institute, Calcutta 700108, INDIA}
\altaffiltext{1}{e-mail: mkafatos@gmu.edu; e-mail: sroy@scs.gmu.edu}
\altaffiltext{1}{e-mail: mroy@scs.gmu.edu}
\altaffiltext{2}{e-mail: sisir@isical.ac.in}
\begin{abstract} 
 Theories of fundamental physics as well as cosmology must 
ultimately not only account for the structure and evolution of the 
universe and the physics of fundamental interactions, but also lead to 
an understanding of why this particular universe follows the physics
 that it does. Such theories must ultimately lead to an understanding of the 
values of the fundamental constants themselves. However, all such efforts have failed, leaving fundamental 
constants outside of any physical theories. In this paper we take
 a different approach than the usual evolutionary picture where 
the physics itself is assumed invariant. We study numerical 
relations among fundamental constants starting from relationships 
first proposed by Weinberg (1972). We have shown (Kafatos et al.2000) that they 
turn out to be equivalent to the relations found by Dirac (1937). Then a new scaling 
hypothesis relating the speed of light $c$ and the scale of the 
universe $R$ is explored. The "coincidences" of Dirac and Eddington(1931) concerning 
large numbers and ratios of fundamental constants do  not need to be explained in our view, 
rather they are accepted as premises and in the process, they yield a fundamentally different 
view of the cosmos. We develop an axiomatic approach and the fundamental constants can be assumed to 
vary and this variation leads to an apparent expansion of the 
universe. Also the variation of constants leads to change in the parameters 
like permittivity and  refractive index of the quantum vacuum. This gives rise to a possibility 
of explaining some of anomalies found in the observations of high redshift quasars. The variations 
of the fundamental constants lead to a 
changing universe,i.e., the number of nucleons varies, etc. The 
increase of the number of nucleons and the redshift of the spectral lines 
appear to be related to the emergence of an arrow 
of time as perceived by an observer in the present universe. Possible 
implications of this new approach in astrophysical domains are discussed.
\end{abstract}
\keywords {Fundamental constants; Large number Hypothesis;   
Scaling hypothesis; Anthropic principle; Nucleons, Redshift}
PACS No. : 31.30Jr, 12.20 Ds Jr, 95.30.Dr
\section{Introduction}
At least several deacdes ago, starting with  Milne(1935,1937) and Dirac(1937), questions 
have been arising from time to time whether the Newtonian gravitational 
constant, $G$, be varying in Cosmological time.
Thus Cosmological consequences of allowing some of these constants of 
nature to change, have been studied to evaluate the effects of time-evolution 
of 'constants' in generalizing frameworks of the general theory of relativity 
with the purpose of allowing them to become space-time variables. Through the Scalar-Tensor theory of gravity  
proposed by Brans-Dicke(1961), the variation of the gravitational 'constant' $G$ has been studied 
extensively. Bounds of a possible variation of the fundamental physical constants at the epoch of primordial 
nucleosynthesis are determined by Ivachik et al.(2001). \\
\indent
Currently, following Bekenstein(1982), Sandvik, Barrow and Magueijo(2002a; 2002b) have 
also developed a theory which describes the space-time variation of the fine structure constant. 
This provides framework for the rigorous study of simulteneous variations of their three dimensional 
counterparts(Forg\'acs et al. 1979; Barrow 1987; Damour 1994; Marciano 1984;
Drinkwater et al. 1998). New observational limits have been stimulated by high-quality 
astronomical data(Drinkwater et al. 1998; Webb et al.1999,2001).
Marciano(1984), Barrow(1987) and  Damour et. al.(1994) have shown in their 
three dimensional subspace theory that ``constants''will vary at the same rate 
as any change which is occuring at the scale lengths of the extra compact dimensions. \\
\indent
Damur et al.(1994) showed that
cosmological variation of $\alpha$ may proceed at different rates at different points in 
space-time. Various functional forms for time variations of $\alpha/G$ have been 
derived using the Kaluza-Klein theory and the assumption of constant masses.
Marciano discussed the self-consistency relations required if there are simultaneous 
variations of different constants in unified gauge theories and examined any possible
 non-monotonic variation in $\alpha$ with $t$, using a running coupling dependence 
of strong, weak and electromagnetic interactions to produce self-consistent predictions 
for the simulteneous variation of more than one coupling or mass ratio. This has been 
discussed in detail by Drinkwater et al.(1998), where the variations of $G$ 
and $\alpha$ 
could be linked by relations of the form $\Delta \alpha/\alpha^2 \ \sim \ \Delta G/G$. 
Considering high energy physics, featuring additional dimensions of space and new dilation 
fields, they have provided motivations for studying variations in the gravitational, strong 
and electroweak coupling constants.(Zlatev,Wang $\&$ Steinhardt 1999; Chiba 1999; Antoniadis $\&$ Quiros 1997). \\
\indent
Theories  unifying gravity and other kinds of interactions, 
such as string theory and $M$ theory where, the existence of  the additional 
compact dimensions of space have been considered(Horavath $\&$ Witten 1996a,b), 
suggest the possibility of spatial and temporal variation of 
physical ``constants'' in the Universe. The currently popular scenarios for M-theory 
(e.g.,Antoniadis et al.1998; Arkani-Hamed, Dimopoulos $\&$ Dvali 1998; Randall $\&$ 1999a,b) 
suggest that  gravitational force needs to be assumed to act in all($>3$) spatial dimensions (the 'bulk') 
whilst all other interactions act only in three-dimensional space (the 'brane'). 
Thus observations of the constancy of three-dimensional non-gravitational 
constants in 3-dimensions(like $\alpha$) could therefore be of {\it crucial importance} 
in testing these theoretical scenarios. \\
\indent
A few interesting theories have recently been proposed , namely, - a kind of {\it fine tuning} 
has been tried to be established between the  variation in the fine 
structure constant $\alpha$ and the possible change  of the light propagation 
speed (Moffat 1993; Albrecht $\&$ Magueijo 1999; Barrow 1999). This kind of 
minimal varying speed of light theories offer possible explanation for the so called 
cosmological  problems: the horizon, flatness, 
cosmological constant, entropy and, homogeneity problems. 
Barrow and Magueijo (1998) tried to show that there 
exists a set of duality transformations between these two representations. \\
\indent
On the other hand, recent observations of astrophysical events at high 
redshifts (Schaefer 2003; Amelio-Camella 1998) can be 
used to place severe limits on the variation of the speed of light itself 
($\Delta c/c$), as well as on the photon mass ($m_\gamma$). Schaefer (2003) presented new limits 
on $\Delta c/c < 6.3 \times10^{-21}$ and the lowest limit on $m_\gamma \sim 4.2\times10^{-44}$ from 
explosive events at high redshifts. Lehnert and Roy (2001) also discussed, from the point of the possible 
effect of fluctuation of permittvity and permeability in vacuum that, photons may be gaining mass, 
as if indeed photons have non-zero masses. Recently, Ranada(2003) proposed that due to variation of 
physical constants, there will be change of permittivity and permeability of quantum vacuum, the effect 
of which will lead to the change of refractive index of the vacuum. In that case, there should be an additive 
effect on the rest mass of photon as well as it can give rise to the shift of the frequency of the photon 
propagating through this kind of vacuum. \\
\indent
In this paper, at first, in section II, we mention some important recent experimental observations as well as theoretical 
developments regarding the variation of the physical constants, occuring in a single way or 
simulteneously. In section III, we discuss the deviation of  numerical relations 
and the concept of scaling. Finally, the possible implications for astrophysical 
observations and cosmology are dealt in section IV. 
\section{Fine Tuning as Implied by Experimental and Cosmological Observations}
\noindent
There are a number of observations which must be applied in any 
cosmological theory that attempts to explain the observed structure 
of the universe. Since we have no understanding of why the constants of 
Nature assume the values they do in our universe, whether they are logically 
independent, or , even whether they are truly constant,  it is 
difficult to realize whether only one fundamental constant, one 
at a time, is varying, or all of them do simulteneously vary i.e., if there is a real
 sense of fine tuning in the kind of variation these constants follow. 
\subsection{Experimental Observations}
\noindent
Generally, the direct laboratory measurements provide interesting constraints on time-varying 
$\alpha$ (Prestage et al. 1995). By comparing the rates of two clocks associated with 
different atoms(H-maser and $Hg^+$) over a 140 - d period, they were able to 
constrain $|{\dot \alpha}/\alpha| \leq 3.7 \times 10^{-14} {\rm yr}^{-1}$ \ 
(i.e.,$ |\Delta \alpha/\alpha| \ \leq \ 1.4 \times 10^{-14}$). But these limits are significantly 
weaker than those derived from geophysics 
and astrophysics because of the billions of years of look back time over which the 
latter two fields can gather data.\\
\indent
An analysis of the observed anomalous abundance of $Sm^{149}$ at 
OKLO-phenomenon - a natural  nuclear fission reactor that 
operated at Gabon, West Africa, $\sim 1.8$ billion years ago, also points to 
this limit of variation of $\alpha$ with time. Shlyakhter (1976), 
 following the nuclear resonance level in the $Sm^{150}$ isotope, put on an upper bound  
 on $|\Delta \alpha/\alpha|$. \\
\indent
Damour and Dyson
(1996) also analyzed this problem with a different approach and found more stringent bounds. They 
concluded that the relative change of $\alpha$ from then to now is in interval, given by
$$-0.9 \times10^{-7} < \frac{\alpha^{\rm Oklo} - \alpha^{\rm now}}{\alpha} < 1.2\times10^{-7}$$
$$-6.7 \times10^{-17}{\rm yr^{-1}} < \frac{\dot \alpha}{\alpha} < 5.0\times10^{-17}{\rm yr^{-1}}$$ 
obtained from the constancy of the $K^{40}$ decay rate (Dyson 1972), comparable to the 
limit derived from Big Bang Nucleosynthesis(BBN) : $|\beta^{\rm BBN} -\beta^{\rm now}/\beta  
< 0.06|$. More recently, Fujii et al.(2001) obtained somewhat tighter constrains 
taking new samples from the Oklo reactor: $\Delta \alpha/\alpha = (-0.04\pm 0.15)\times10^{-7}$.
 However, the Oklo limit corresponds to variations at very low ``redshift'' , $z \sim 0.1$ , i.e., in local 
or in a non-cosmological environment. There are several other studies which set bounds on the variation 
of $|\Delta \alpha|$, using a number of different data (Uzan 2002). \\
Experimentally, quasar (QSO) absorption lines, and particularly the  detection of high-redshift 
absorption systems which are intersecting the lines of sight 
towards distant qausars provide an ideal and powerful tool in a cosmological 
settings where one can search  for possible temporal or even spatial variations in 
the assumed fundamental constants of Nature. \\
\indent
Savedoff(1956) first analyzed doublet separations seen in galaxy emission spectra 
to obtain constraints on the variation of the most observationally sensitive constant, namely the  
electromagnetic fine structure constant $\alpha = e^2/\hbar c$. Various propositions 
and ideas, since 1930, together with the first constraints from spectroscopy of QSO 
absorption systems, starting from the 1960s, are given in detail by Varshalovich \& Potekhin (1995).  
Tight constraints on $\Delta \alpha/\alpha$ come from optical absorption-line studies. 
Drinkwater et al(1998) and Carilli et al(1998) considered the bounds that can be placed 
on the variation of the fine structure constant and proton $g$ factor from radio observations 
of atomic and molecular transitions in high-redshift quasars which have further been 
constrained to smaller values for $\alpha$ at higher redshifts. \\
Observations of Webb et al.(2001a,b) confirmed these results with improved 
techniques and extended previous results to a higher-redshift sample of damped Lyman-$\alpha$ 
systems. They studied relativistic transitions to different ground states using absorption lines QSO spectra 
by exploiting  the extra sensitivity of  many-multiplet technique. 
The trend of all these results appears to be that the value of $\alpha$ {\it was lower in the past}, 
with $\Delta \alpha/\alpha = -0.72 \pm 0.18 \times 10^{-5}$ over $z \approx 0.5 -3.5$
(spanning $\sim 23\%$ to $87\%$ of the age of the universe).  \\
\indent
The
 most precise constraint to date was obtained by Murphy et al.(2001c) i.e., 
$\Delta \alpha/\alpha = (-0.5 \pm 1.3) \times 10^{-5}$, by analysing 21 SiIV doublets 
$(2<z<3)$ in 13 QSO spectra which thus  provide strong evidence that the fine 
structure constant might be changing with cosmological time (Murphy et al.2001a,b,c; Webb et al.1999;2001). 
They also considered the implications of simulteneous variations of several ``constants'' and 
showed how these observational limits can be used to constrain a class of inflationary universe 
theories in which small fluctuations in the fine-structure constant are also predicted to occur. \\ 
Other investigations (Avelino et al.2000, 2001; Battye et al.2001) have claimed preferred 
nonzero values of $\Delta \alpha/\alpha <0$ to best fit the cosmic microwave background (CMB)
and Big Bang Nucleosynthesis (BBN) data at $z \approx 10^3$ and $z \approx 10^{10}$, 
respectively, but result in much larger variations. \\
\indent
Another group (Varshalovich et al. 2001) studied this problem of possible variation of the fundemental 
physical constants at the epoch of quasar spectra formation (i.e.,$\sim 10$ billion years ago). They 
calculated the upper limits of this variation basing  on an analysis of absorption spectra of Quasars 
with high redshifts and also applied a number of systemic efects, which can simulate variation of the 
constants.   
\subsection{Cosmological Observations}
\noindent
We now turn our attention to several cosmological observations and discuss some of 
their implications on the nature of the universe. \\
\noindent
{\bf (a)} The universe appears to be quite flat, in other words the density of 
the universe is very close to the so-called closure or critical density,
\begin{equation}
\rho_{\rm crit} = 2 \times 10^{-29}(\frac{H_0}{100 {\rm km s}^{-1} {\rm Mpc}^{-1}})^2 
{\rm gr \ cm^{-3}}
\end{equation}
\noindent
where $H_0$ is the Hubble constant defined as the apparent rate of expansion 
with distance, ${\dot R}/R$  and  $R$ being the scale of the universe. The observed density is not really 
equal to the closure density when one observes regular, luminous matter. In big 
bang cosmology, the so-called "Hubble constant" is actually a function of cosmic 
time, i.e., it is a variable. Its present-day value seems to be $\sim$ 
$75 {\rm km}{\rm s}^{-1}{\rm Mpc}^{-1}$. The universe appears to be close (but still 
off by factor of $\sim 10 - 100$ from the closure limit, at present) to 
a flat, Euclidean, Einstein-de Sitter state as indicated from (1), 
and yet it is still not clear what the geometry of the universe is,i.e., whether 
exactly flat (which would be required by the inflationary scenario); 
open (yielding a forever-expanding, negatively curved space-time); 
or closed (yielding a maximum expansion and a  positively curved space-time). \\ 
\noindent
{\bf (b)} If one is to assume that the universe followed an inflationary period in 
the distant past, then the universe must have been exactly flat to one part in 
$10^{50}$ near the time of Big Bang. This is so-called {\it flatness} problem: 
This is such a remarkable requirement that the usual interpretation proposed 
in the early $80$'s was that -- early on, the universe
 was in an inflationary state, washing out any departures from flatness on time
 scales of $10^{-35}$sec. The inflationary model proposed by Guth(1981) and 
others has been developed in various forms to account for the flatness of the 
universe and also is proposed to solve the horizon problem, or apparent homogeneity of 
the $2.73 {\rm K}$ black body radiation seen by COBE (Smoot 1996). The latter 
problem involves the observation that although the $2.73 {\rm K }$ radiation was 
emitted $\sim 10^5$ years after the beginning, opposite sides of the sky at 
that time were out of causal contact, seperated by $\sim 10^{7}$ light years. 
Other structures involving large-scale correlations in the universe exist such 
as very large structures in the distribution of matter (Geller et al., 1989). 
These structures may be progressively hierarchical all the way to the scale of 
the universe itself. \\
\noindent
{\bf (c)} If the universe is indeed flat, observations indicate that baryons (or 
luminous matter) can only contribute at most $\sim 0.05$ of the closure 
density at present. We should ultimately be able to detect the other $90\%$ 
or more of 
the matter required to give closure density, presumed to be in the form of 
cold dark matter (Novikov, 1996). Nevertheless, attempts to detect such exotic 
matter in the laboratory have, so far, failed. Moreover, the recent 
realization that the cosmological constant $\Lambda$ may have to be 
re-introduced (Peebles, 1998)  to account for the possibility of an accelerating 
universe, has also led to the probability of $\Lambda$ 
itself varying and other similar notions (Glanz, 1998). Barrow $\&$ 
Magueijo (2000) developed a particular theory for varying $c$ (or $\alpha$) 
in which the stress contributed by the cosmological constant varies through 
the combination $\Lambda c^2$. They also showed how the observed non-zero 
cosmological acceleration (Schmidt et al.1998, Perlmutter et al. 1999) 
might be linked to a varying $\alpha$. The case of varying $c$ theories 
are considered to be driven by a scalar field , coupled to the gravitational effect of pressure. The 
very slow variation of the scalar field makes possible for slow variation of $c$ 
which at the radiation era converted the $\Lambda$ energy density into radiation, 
thus preventing $\Lambda$ dominance;  but at the pressureless matter era the situation reversed. \\
\indent
This kind of theory allows variations of $c$ or $\alpha$ to be $\sim 10^{-5}H_0$ 
at $z\sim1$ and yet the associated $\Lambda$  term can be dominant today and 
produce the much needed acceleration. Inflationary universe models provide a possible 
theoretical explanation for proximity to flatness but no explanation for the 
smallness of the cosmological constant itself. Nevertheless, without some direct  
laboratory verification or overwhelming requirements imposed by 
particle theory (neither of which presently exists), the nature of dark matter 
remains elusive. This is clearly a very unsatisfying situation.  \\
\noindent
{\bf (d)} As we saw, present-day approximate flatness yields to an exact flatness in 
the distant past (this was one of the main reasons why the inflationary 
scenario was introduced to begin with). The alternative is to accept {\it fine 
tuning} in the universe. 
In fact, the flatness of the universe is not the only 
fine tuning. In considering other fundamental observed facts, the universe 
appears to be extremely fined tuned. It was Eddington (1931, 1939) and Dirac 
(1937) who noticed that certain cosmic "coincidences" occur in nature linking 
microscopic with macroscopic quantities (Kafatos, 1989). 
A most unusual relationship is the ratio of the electric force to gravitational force 
(this ratio is presumably a constant in an expanding universe where the 
physics remains constant), or
\begin{equation}
\frac{e^2}{G m_e m_p}  \sim 10^{40}
\end{equation}
while the ratio of the observable size of the universe to the size of an 
elementary particle is, or 
\begin{equation}
\frac{R}{(\frac{e^2}{m_e c^2})}  \sim 10^{40}
\end{equation}
\noindent
Here, in this relationship, the numerator is changing as the universe 
expands because the scale of the universe $R$ is constantly changing in an 
expanding universe. \\  
Dirac formulated the so-called {\it Large Number Hypothesis} which simply 
states that the two ratios in (2) and (3) are in fact equal for all practical 
purposes and postulates that this is not a mere coincidence. Various attempts 
were made to account for the apparent equality: a possibility that constants 
such as the gravitational constant $G$ may be varying was proposed by Dirac
(1937) himself and others (Dyson, 1972). Other ratios such as the ratio of 
the size associated to an elementary particle, like the electron, to the Planck length, 
\begin{equation}
\frac{(\frac{e^2}{m_e c^2})}{(\hbar G/c^3)^\frac{1}{2}} \ \sim \ 10^{20}
\end{equation}
\noindent
can also be constructed (Harrison, 1981) yielding to the conclusion that fine 
tuning is prevalent in this universe. These relationships may be indicating 
the existence 
of some deep, underlying harmonies involving the fundamental constants and 
linking the microcosm to the macrocosm. Physical theory has not, however, 
accounted for these in a self-consistent way, waiting perhaps for the 
anticipated unification of all physical forces at the quantum gravity or 
superstring levels. \\
\noindent
{\bf (e)} Other, less traditional ways, such as the Anthropic Principles 
 emerged from attempts by Whitrow (1955; Barrow $\&$ Tipler, 1986)
to understand why it is not surprising that we find space to have three dimensions, and by
 Dicke(1957;1961) to understand the inevitability of Dirac's ''Large number'' 
coincidences in cosmology for the above fine tuning properties of the universe which
 provides some novel anthropic perspectives on the evolution of our universe 
or others. There have been many investigations of the apparent, might be termed 
``finely tuned'', coincidences that allow complexity to exist in the universe(
Carr $\&$ Rees 1979; Tegmark 1998; Hogan 2000). \\
\indent
Recently, a phenomenological and Newtonian 
model has been proposed by Ranada(2003) to explain the recent observed 
cosmological variations of the fine structure constant as an effect of the quantum 
vacuum. He  assumes a flat universe with cosmological constant $\Lambda$  in the 
cases ($\Omega_M,\Omega_\Lambda$) equal to $(0.3, 0.7) $(Perlmutter 1999) and $(1,0)$ respectively. This model 
predicts, that $\Delta \alpha/\alpha$ is proportional to 
$(\Omega_M[R(t)^{-1}] - 2 \Omega_\Lambda[R(t)^2 - 1])$, $R(t)$ being the scale factor 
and shows some kind of agreement with the observations (Webb et al. 2001); however, 
limitations at the present state of developement of his theory  remain.
\section{Numerical Relations and Concept of Scaling}
\noindent
The critical density of the universe in (1) is defined as
\begin{equation}
\rho_{\rm crit} = \frac{3 H_0^2}{8 \pi G}
\end{equation}
\noindent
Let $N_p$ be the number of nucleons in the universe, then writing the mass of a 
particle in terms of cosmological quantities, we have
\begin{equation}
m_p \ = \ \frac{M}{N_p} = \frac{R {\dot R}^2}{2 G N_p}
\end{equation}
\noindent
where $m_p$ and $M$ are the mass of the nucleon and mass of the universe, 
respectively. \\
Weinberg(1972), on the otherhand,  noticed that one can find a relationship linking 
the masses of elementary particles, such as pions, to the Hubble constant and other 
fundamental constants; for example,
$$m_\pi \ \sim \ (\frac{8 \hbar^2 H_0}{G c})^{\frac{1}{3}} \ \ \ \ {\rm and} \ \ \ \ \ m_e \ \sim \ 
(\frac{\hbar e^2 H_0}{(8 \pi)^3 G c^2})^\frac{1}{3}$$
\noindent
where, $m_\pi$ and $m_e$ are the pion and electron masses, respectively.
These relations can be rewritten as 
\begin{equation}
m_p \ \sim \ \chi_{p \pi}(\frac{8 \hbar^2 (\frac{\dot R}{R})}
{G c})^{\frac{1}{3}} \ \ \ \ {\rm with} \ \ \ \ \chi_{p \pi}  =   \frac{m_p}{m_\pi}
\end{equation}
\begin{equation}
m_p \ \sim \ \chi_{pe}(\frac{\hbar e^2( \frac{\dot R}{R})}{G c^2(8 \pi)^3})^
\frac{1}{3} \ \ \ \ {\rm with} \ \ \ \ \chi_{pe}  =  \frac{m_p}{m_e}
\end{equation}
\noindent
From equation(6) and above one can easily get 
\begin{equation}
G^2 \hbar^2 c^{-1} \ \sim \ 
\chi_{p \pi}^{-3}N_p^{-3} \frac{R^4 {\dot R}^5}{64}
\end{equation}
and 
\begin{equation}
m_p = \chi_{p_*}\sqrt{\frac{\hbar c}{G}}
\end{equation}
for, $\chi_{p _*} = \frac{m_p}{m_*}$, and  \ $m_*$ being the Planck mass. Suffix $*$ indicates 
in general Planck quantities. Combining (10) and (6), yields
\begin{equation}
c G \hbar \ \sim \ \frac{1}{4}N^{-2}_p {\chi}^{-2}_{p_*}R^2\dot R^4
\end{equation}
Similarly from (9) and (10), we can have
\begin{equation}
c \ \sim \ 2^\frac{2}{3}N^{-\frac{1}{3}}_p \ \chi^{-\frac{4}{3}}_{p_*}
\chi_{p\pi}\dot R
\end{equation}
The multipling factor for $\dot R$ in (12) is of the order unity, or  
$$ 2^\frac{2}{3}N^{-\frac{1}{3}}_p \chi^{-\frac{4}{3}}_{p_*}
\chi_{p\pi} \ \sim \ 1 $$
\noindent
Conversely, if we choose to set the required condition 
$ 2^\frac{2}{3}N^{-\frac{1}{3}}_p \chi^{-\frac{4}{3}}_{p_*}\chi_{p\pi} = 1$,
  one gets the simple relationship linking the speed of light to $\dot R$, 
i.e., $c = \dot R$ with $N_p \ \sim 3.7 \times 10^{79}$, which is a good 
estimate of the number of particles in the current universe. The relationship 
$c = \dot R$ could be intrerpreted as the Hubble Law \ $\dot R \ \sim c$, 
although we emphasize that this is just a relationship and might not imply 
that an expansion is indeed taking place. We can arrive at the similar 
conclusion if one works with the relations using electrons. Now, if we start by assuming a heuristic relation
$$c\equiv \dot R$$
i.e., the speed of light is identical to the rate of change of the scale 
of the universe, we can construct an axiomatic approach equivalent to the Hubble Law. 
This axiomatic approach can be considered as an alternative approach to the 
mysterious coincidences of Eddington and Dirac which Weinberg called "so 
far unexplained... a real, though mysterious significance."  \\
It can be further shown that all lengths, such as the Planck length, $l_*$, 
the classical electron radius, $r_e$, etc., are all proportional to the scale of the universe, i.e., 
\begin{equation}
l_* , r_e,  \sim \ (\cdots)R
\end{equation}
For example, 
$$l_*  \ \sim (2^{-\frac{7}{3}}N^{-\frac{1}{3}}\chi^{\frac{5}{3}}_{p_*}
\kappa^{-2}_{p\pi})R$$

Similar relations can be formed for $r_e$ and $r_p$ where $r_e$ and $r_p$ 
are the electron and proton radii respectively. From (11) and (13) we obtain
\begin{equation}
G \hbar = \frac{R^2 {\dot R}^3}{4} N_p^{-2}\chi^{-2}_{p_*} \ 
\sim \ 3.4 \times 10^{-122}R^2 {\dot R}^3
\end{equation}
\noindent
a relationship linking the gravitational and Planck's constant to $R$ 
and $\dot R$ and where the last relationship (14) holds for the current values of 
 $ N_p^{-2}\chi^{-2}_{p_*}$.  Let us now set the following initial conditions, i.e., \ \ $R \ \rightarrow l_* \ \ \ {\rm and} \ \
 \dot R \ \rightarrow \frac{l_*}{t_*}$. Here $l_*$ and $t_*$ are the Planck length and Planck time respectively. Then 
$$ N_p^{-2}\chi^{-2}_{p*}/4 \ \rightarrow 1  \ \ \ \ \ \ {\rm \ at \ those \ initial \ conditions}$$
$$ N_p^{-2}\chi^{-2}_{p_*}/4 \ \sim 3.4 \times 10^{-122} \ \ {\rm \ for \ the \ present \ universe}$$
\noindent
The limit $N_p \rightarrow 1$ indicates that in our model `{\it in the beginning} 
there was only one bubble-like object or a  
{\it cosmic egg} (Israelit $\&$ Rosen 1989). Moreover, 
$R \rightarrow l_*$ and $N_p \rightarrow 1$ imply that 
$\chi_{p_*} \rightarrow 1$ as well (similarly for all ratios of 
masses $\chi$'s), which in turn indicates that the masses of all particles 
were equal to each other at these initial conditions. \\
In the beginning,
$$\frac{R}{(e^2/m_ec^2)} \ \sim \ \frac{(e^2/m_ec^2)}{(G/c^3)} \ \sim \ 1$$ 
rather than the large values of $10^{40}$ and $10^{20}$ which these ratios are 
equal to, respectively, today and also, all lengths were equal, all 
masses were equal and there was only one particle or {\it cosmic egg}. Today, these 
ratios are not unity, as there is a very large number of particles in the universe 
and $R$ is equal to $\sim 10^{28}$ cm. However, scale-invariant relationships 
such as  $c \equiv \dot R$, all lengths are proportional to each other, etc. 
still hold. \\
\indent
Israelit and Rosen (1989) proposed a cosmological model where the 
universe emerges from a small bubble ({\it cosmic egg}) at the bounce point of a 
de Sitter model filled with a cosmic substrate ({\it prematter}).
In other words, $c \equiv \dot R$, at the {\it initial time} 
when $N_p \rightarrow 1$  and all $\chi \rightarrow 1$, and this relationship 
remains invariant even at the present universe (cf. equations (12) and (13)). The 
self-consistency is obtained by calculations for the value of $N_p$ from (12) 
and (14). This relation is a type of a scaling law and connects the microcosm 
to the macrocosm. \\
\indent
Now, if irrespective of the presence or absence of expansion of 
the universe, $R$ itself is changing from the Planck scale to the size of 
the observable universe, then the fundamental constants like $G,\hbar$ and $c$  
are changing simultaneously. \\
\indent
Note, however, that we cannot deduce the actual 
variation or the initial value of $c$ and other constants from observations: 
The relationship $c \ \equiv \ \dot R$  
is not enough to tell us the actual variation or even over {\it how long} it takes 
place. It is a scale invariant relationship. If we re-write it as a 
scale-invariant relationship, 
$$\frac{c(t_*)}{c(t_0)} = \frac{\dot R(t_*)}{\dot R(t_0)}$$ 
where $ t_*$ and  $t_0$ could be conveniently taken as the Planck time and the 
present {\it age} of the universe, then this relationship is not enough to give us 
the evolution of  $\dot R$ or even the values of $t_*$ and $t_0$. \\
\indent
Hence it cannot tell us 
how $c$ itself is varying or even if it is varying. If we wanted to insist that $c$ is 
{\it constant}, then all the other "constants" like $G$ and $\hbar$ {\it are really constant 
as well}. But if $c$ is not constant, then all the other "constants" are varying as well. 
In both cases, however, the number of particles is changing, the ratios of 
masses are changing and the ratios of scales or lengths are also changing. 
An arrow of time {\it could}, therefore, be introduced. In this picture, invariant 
relationships hold and from unity, there is evolution into diversity. One 
cannot, though, conclude how the variations are taking place, over what timescales 
they are taking place or even how old the universe is. The universe could be $10^{10}$ 
years old or $5 \times 10^{-44}$ sec (the Planck time) old, or any time in between. {\it Time is 
strictly a parameter} that can be introduced in the scale-invariant relationships. 
It has no meaning by itself. The universe appears to be evolving as the number of 
particles and ratios are varying.
\section{Implications in the Astrophysical domain}
\noindent
The variation of physical constants may change the permittivity and permeability of
the underlying quantum vacuum which play significant role in the astronomical domain.
  According to the phenomenological Newtonian model, presented by Ranada (2003), the cosmological 
variation of the fine structure constant is due to the combined effect of the fourth Heisenberg 
relation and the gravitational interaction of the virtual pairs in the zero-point radiation with 
all the universe. More precisely, it is argued that, because of the fourth Heisenberg relation, the 
density of the sea of virtual particles in the quantum vacuum must change in a gravitational field, 
with a corresponding variation of permittivity and permeability that depends on the average 
gravitational potential of the universe ($ \phi$). In his model, the quantum vacuum is 
treated as a transparent optical medium charaterized by its permittivity. 
As a result, the contribution due 
to change of $\alpha$ in the frequency shifts of the spectral lines is  
$$\frac{\triangle\omega}{\omega} = 4 \beta^\prime \frac{\triangle\phi}{c^2}$$
$\beta^\prime$ being certain coefficient.
Though they might appear as similar, this is different from the redshift expected due to 
gravitational redshift which is
$$\frac{\triangle\omega}{\omega} = \frac{\triangle \phi}{c^2}$$
$\phi$ being the Newtonian gravitational potential.
Hence total redshift,expected to be observed is due to these two combined effetcs, given by,
$$ \frac{\triangle\omega}{\omega} = (1+4\beta^\prime)\frac{\triangle\phi}{c^2}$$
\indent
 However, Ranada pointed out a necessary condition for the compatibility 
of his results and that of gravitational redshift experiments as  $ \xi\neq 4\times10^{-3}$ , $\xi$ being a
a parameter related to renormalization effects of the quantum vacuum. It is interesting to note that 
according to this model, light is also effected by the gravitational potential $\phi$ so that 
it was slower in the past and the optical density of the quantum vacuum increases 
towards the past and decreases as the universe ages. However, he had to put a boundary 
on the value of $\beta$ and consequently on  $\xi$ in order to make his results compatible 
with that of gravitational redshift experiments i.e., 
\ $\xi \sim 1.3\times10^{-5}$ for his model and \ $1.9\times10^{-5}$ for taking the two cases 
togather. The best confirmation of the gravitational redshift, 
those of Pound, Rebeka and Snider(Weinberg 1973) agree with the prediction of General Relativity up to about 
1\% , but they also refer to nuclear levels in which the electromagnetism plays a part. \\
\noindent
We think this incompatibility is due to the fact that he did not consider the added 
effect due to non-zero photon mass. One of the main difficulties with all these 
approaches is that no change in width of the spectral lines have been observed. 
This is contrary to the present astronomical findings, especially, in quasar
astronomy where one can get many broaddened lines as the redshift becomes higher. \\
\indent
 Lehnert and Roy(1998) 
showed that if the light is propagated through Maxwell vacuum with different permittivity, permability and
refractive index than the ususal vacuum, then the photon will loose its energy and there will be a shifting
in the spectral lines as well as the photon gaining mass. From the estimated values of permittivity and permeability 
one can estimate the lower bound of non-zero rest mass of the photon (Kar, Sinha $\&$ Roy 1993;1996) whose presence 
could be manifested in laboratory experiments. It is also evident now, from the above discussions, that the redshift due 
to these three effects is an indication of arrow of time to an observer within the universe. \\
\indent
According to Lehnert and Roy, Maxwell's equations in 
vacuum is  modified by assigning a small nonzero conductivity($\sigma$).
As a first step, we can extend  modified Maxwell's equations, assigning  a nonzero space-charge {\it in vacuo} 
to it. Then, if a nonzero conductivity coefficient is assigned to this Maxwell vacuum instead of space-charge 
then the photon looses its energy when it propagates through such a vacuum. At first, let us consider the Maxwell equations 
with $\sigma \neq 0$, i.e., 
$${\rm div}{\bf E} = 0 \ \ \ \ \ \ \ {\rm curl}{\bf H} = \sigma {\bf E} + \epsilon_0\chi_e\frac{\partial {\bf E}}{\partial t}$$
$${\rm div}{\bf H} = 0 \ \ \ \ \ \ \ {\rm curl}{\bf E} = \mu_0\chi_m\frac{\partial {\bf H}}{\partial t}$$
\noindent
where, \\
\noindent
$\mu_0$ = the vacuum permeability constant \\
$\chi_e$ = the relative dielectric constant \\
$\chi_m$ = the relative permeability constant. \\
\noindent
Here, the four current is given by
$$j = ({\bf j},j_0) \ \ \ \  {\rm with} \ \ \ \ {\bf j} =\sigma{\bf E}; \ \ \ j_0 = 0$$
Again,
$$\bigtriangledown \times\bigtriangledown\times{\bf E} = -\bigtriangledown^2{\bf E}$$
which together with Maxwell's equations gives
$$\bigtriangledown^2{\bf E} = - \frac{\epsilon_0\chi_e\chi_m}{c^2}\mu_0\frac{\partial^2{\bf E}}{\partial t^2} 
+ \sigma\mu_0\chi_m\frac{\partial {\bf E}}{\partial t}$$
\noindent
This equation is not time invariant. The second term on the right hand side indicates that there 
will be a dissipation of energy during the propagation of a photon. 
Considering a plane wave in the z-direction,
$$E_x = be^{i\omega(t-z/v)}$$
$$H_y = b\left (\frac{\epsilon_0\chi_e}{\mu_0\chi_m} \right)^{\frac{1}{2}} {\rm exp}i\omega(t-\frac{z}{v})$$
and putting $(q = \frac{1}{v})$, we get
$$q^2 = \frac{\epsilon_0\chi_e\chi_m}{c^2} \left (1 - \frac{i \sigma}{\omega \epsilon \chi_e} \right )$$
\noindent
The velocity defined by $v$ above will give rise to a complex refractive index in vacuum. The real part 
of $q^2$ gives rise to a phase velocity of propagation of the disturbance through the underlying vacuum. 
Taking the real and imginary part as $\alpha$  and $\beta$  respectively, $E_x$ and $H_y$  can be 
shown to be proportional to 
$${\rm exp} (-\omega \beta z){\rm exp}(t - \alpha z)$$
and the complex quantity $q$ can be written as
$$q = \alpha - i\beta$$
with
$$ \alpha^2 = \frac{\chi_e\chi_m}{2c^2}\left [\left (1 + (\frac{\sigma}{\epsilon_0\chi_e \omega})^2\right)^{\frac{1}{2}} + 1 \right ] $$
$$\beta^2 = \frac{\chi_e\chi_m}{2c^2}\left [ \left (1 + (\frac{\sigma}{\epsilon_0\chi_e \omega})^2 \right )^{\frac{1}{2}} - 1 \right ] $$
\noindent
Then the following situation arise : \\
\noindent
{\bf (a)} Plane waves are progressively damped with the factor ${\rm exp}(-kz)$, where $k=\omega\beta$. \\
{\bf (b)} The phase velocity of propagation of the wave is $1/\alpha$  and varies with the frequency. \\
\indent
In the limit $\frac{\sigma}{\omega} \rightarrow 0$, we have
$$\alpha \simeq 1 + \frac{1}{8} \left ( \frac{\sigma^2}{\epsilon_0^2\chi_e^2}.\frac{1}{\omega^2} \right ) + {\cal O}(\frac{\sigma^4}{\omega^4}); \  \ \ \ \ 
\beta^2 \simeq \frac{1}{2}.\frac{\sigma^2}{(\epsilon_0\chi_e)^2}.\frac{1}{\omega^2}$$
\noindent
Then the phase velocity $v_p$ and the group velocity $v_g$ of propagation of the disturbance through the underlying vacuum, after some 
calculations becomes
$$v_p = \frac{c}{(\chi_e\chi_m)^{\frac{1}{2}}}\left (1 - \frac{1}{8}. \frac{\sigma^2}{(\epsilon_0\chi_e)^2}. 
\frac{1}{\omega^2} \right )$$
$$v_g = \frac{c}{(\chi_e\chi_m)^{\frac{1}{2}}} \left [\sqrt{1+\frac{1}{4}\frac{\sigma^2}{(\epsilon_0\chi_e)^2}.\frac{1}{\omega^2}} \right] $$
\noindent
However, in the limiting case, $\sigma = 0$, we have \ \ $v_p = v_g = c$. \\
\noindent
Now taking $v_g$ as the velocity of photon and $m_\gamma$ as the nonzero mass of photon, we have
$$ E = h\nu = \frac{m_\gamma c^2}{\sqrt{1 - \frac{v_g^2}{c^2}}}$$
and the mass of photon becomes
$$m_\gamma^2 = \frac{h^2\nu^2}{n^2c^4} \left [(n^2 - 1)-\frac{\sigma^2}{(\epsilon_0\chi_e)^2}\frac{1}{\omega^2} \right ] 
\ \ \ {\rm for} \ \ \ n =\sqrt{\chi_e\chi_m}$$
\noindent
But this is unphysical. But if we instead introduce the phase velocity in the de Broglie relation, 
we get a physical solution i.e., a real nonzero rest mass of the photon. Finally, we 
get for $n\sim 1$,
$$m_\gamma \simeq \frac{\sigma h}{\sqrt{2(\epsilon_0\chi_e)}}.\frac{\pi}{c^2}$$
\noindent
Let us now consider the variation of permittivity($\epsilon$)and permeability ($\mu$), following Ranada (2003) in addition to the 
above mentioned effect. Then expressing the relative permittivity and permeability at a space time point with 
a weak gravitational potential $\phi$, we get,
$$\epsilon_r = 1 - \beta^\prime(\phi - \phi_\oplus)/c^2, \ \ \ \ \mu_r = 1 - \gamma^\prime(\phi - \phi_\oplus)/c^2$$
\noindent
$\beta^\prime$ and $\gamma^\prime$ being certain coefficients, which must be positive since quantum vacuum 
is dielectric but paramagnetic. $\phi_\oplus$ here, represents the present gravitational potential of all 
the universe at earth. Finally, taking into consideration all these variations together, we can write the velocity of light, changed to 

$$c^\prime \simeq \frac{c}{\sqrt{\epsilon_r\mu_r}} = c\left [1+(\beta^\prime + \gamma^\prime)(\phi-\phi_\oplus)/2c^2 \right] = c^2 . \bar{m_\gamma}$$ 
where,
$$\bar {m_\gamma} = m_\gamma \left [ 1 + (\beta^\prime + \gamma^\prime) \frac{\phi \phi_\oplus}{2c^2} \right ]$$ 
\noindent
Thus, the effective non-zero photon, gaining mass as a result of non-zero 
conductivity-coefficient in vacuum, calculated by Lehnert and Roy, will be 
modified further due to the variations of the physical constants, if any. It is interesting to note that in such cases, 
i.e., if there is variation of physical constants due to the variation of permittivity and permeability then, 
one should get redshift due to non-zero rest mass of photon too, in addition to the above effects.
\section{Discussion and Conclusions}
\indent
The existence of horizons of knowledge in cosmology, indicate that as a horizon 
is approached, ambiguity as to a unique view of the universe sets in. It was 
precisely these circumstances that apply at the quantum level, requiring that 
complementary constructs be employed (Bohr 1961). At the {\it initial time}, which 
could be conveniently taken as the Planck time, if we set the conditions 
like $c \equiv\dot R$, as proposed in this paper, we can axiomatize the numerical 
relations connecting the microcosm and the macrocosm. One then has scale-invariant 
relationships. During the {\it evolutionary} process of the universe, the fundamental 
constants are changing or they may be constant. In the former case, we don't 
know how or even over what timescales they are changing. In the latter case, 
one gets the usual evolutionary universe. This is a clear case where 
complementarity applies. \\
\indent
In other words, as $N_p$ is changing from the initial value of $1$ (unity) to the 
present large value of $\sim 10^{80}$ (diversity), more particles are created 
as $R$ and all length scales as well as all masses are changing. 
This could be interpreted by an observer as an "expansion of the universe". An 
observer, who is inside the universe will perceive an "arrow of time" and an 
"evolving universe". But equivalently, as the "constants" change ( they would {\it all} 
have to be changing), there appears to be an evolution. As $N_p \ \rightarrow 10^{80}$, the 
present number of the nucleons in the universe, the fundamental "constants" 
achieve their present values. \\
\indent
To recapitulate, the arrow of time can be related to a kind of complementarity 
between two constructs, i.e., the fundamental "constants" {\it are truly constant}, on 
the one hand; and the fundamental "constants"  {\it are changing}, on the other hand. \\
\indent
In summary, we found that by adopting Weinberg's relationship (which can be shown 
to be equivalent to Dirac's relationships (2) and (3) when the latter are equated 
to each other), we can obtain a relationship linking the speed of light $c$ to the 
rate of change of the scale of the universe. In fact, the proportionality factor 
is $\sim 1$ if one substitutes for values of fundamental quantities like the 
present number of particles in the universe, etc. The next step assumes that the 
relationship linking $c$ and $R$ is an identity, i.e. $c \equiv \ \dot R$   for 
example, at the Planck time, one observes that this relationship still holds if 
the ratios of all masses $\rightarrow 1$ and the number of particles 
also $\rightarrow 1$. As such, it is possible (but not necessary) to state that 
{\it all} the fundamental constants are changing and not just one of them as was 
assumed in past works. It is interesting that, recently, the possibility of 
the cosmological constant $\Lambda$ itself changing (Glanz, 1998, Perlmutter 1999 $\&$ 
references there in) has been suggested. As such, 
what we are suggesting here as a framework for the universe is -- a natural 
extension of previous ideas. Therefore, as $N_p$ changes from an initial value 
of $1$ to the present value of $10^{80} (1 \rightarrow 10^{80})$, the universe 
would be appearing to be evolving to an observer inside it or an arrow of time 
would be introduced.  \\
\indent
Again due to the variation of physical constants, the structure of quantum vacuum will also be 
changed as a result of which there will be a redshift as an effect of changing permittivity and 
permeability of the vacuum. The evidence of this kind of redshift can be related to the arrow of 
time will appear as an arrow of time to an observer within the universe similar to that due to 
change of number of nucleons. Finally, the outcomes of this prescription are not just that an 
arrow of time is introduced and the mysterious coincidences of Dirac and Eddington now can be understood as 
scale-invariant relationships linking the microcosm to the macrocosm; in addition, all scales are 
linked to each other and what one calls, e.g. the {\it fine structure constant, fundamental length}, 
etc. are  purely a convention and interrelated. In the same way, time itself is not as fundamental as 
the scale-invariant relationships linking the microcosm to the macrocosm. 
 
\acknowledgements
\indent

The authors (S.Roy and M.Roy) greatly acknowledge the hospitality and funding provided by the School of computational 
Sciences and Center for the Earth Observing and Space Science, George Mason University, USA for this work.

\end{document}